\documentclass[prd, aps, superscriptaddress, preprintnumbers, twocolumn, floatfix, nofootinbib]{revtex4}

\usepackage{amsfonts}
\usepackage{amsmath}
\usepackage{amssymb}
\usepackage{bm}
\usepackage{dcolumn}
\usepackage{epsfig}
\usepackage{graphicx}   
\usepackage{graphics}
\usepackage{color}
\usepackage[latin1]{inputenc}
\usepackage{latexsym}
\usepackage{rotating}
\usepackage{hyperref}

\usepackage{amsfonts}
\usepackage{amsmath}
\usepackage{amssymb}
\usepackage{bm}
\usepackage{dcolumn}
\usepackage{epsfig}
\usepackage{graphicx}
\usepackage{graphics}
\usepackage[latin1]{inputenc}
\usepackage{latexsym}
\usepackage{rotating}
\usepackage{hyperref}

\newcommand\be{\begin{equation}}
\newcommand\ba{\begin{eqnarray}}
\newcommand\ee{\end{equation}}
\newcommand\ea{\end{eqnarray}}
\newcommand\lmk{\left(}
\newcommand\rmk{\right)}

\begin{document}

\title{Note on Reheating in G-inflation}

\author{Hossein Bazrafshan Moghaddam}
\email{bazrafshan@physics.mcgill.ca}
\affiliation{Department of Physics, McGill University,
Montreal, QC, H3A 2T8, Canada} 

\author{Robert Brandenberger}
\email{rhb@physics.mcgill.ca}
\affiliation{Physics Department, McGill University, Montreal, QC, H3A 2T8, Canada, and \\Institute for Theoretical Studies,
ETH Z\"urich, CH-8092 Z\"urich, Switzerland}

\author{Jun'ichi Yokoyama}
\email{yokoyama@resceu.s.u-tokyo.ac.jp}
\affiliation{Research Center for the Early Universe (RESCEU), Graduate School of Science, The University of Tokyo, Tokyo 113-0033, Japan, \\
Department of Physics, Graduate School of Science, The University of Tokyo, Tokyo, 113-0033, Japan\\
Kavli Institute for the Physics and Mathematics of the Universe (Kavli IPMU),
UTIAS, WPI, The University of Tokyo, Kashiwa, Chiba, 277-8568, Japan}

\date{\today}

\begin{abstract}

We study particle production at the end of inflation in kinetically driven G-inflation model and
show that, in spite of the fact that there are no inflaton oscillations and hence no
parametric resonance instabilities, the production of matter particles due to a
coupling to the evolving inflaton field can be more efficient than pure gravitational
Parker particle production.

\end{abstract}

\maketitle

\section{Introduction} 

{\it Reheating}, namely the transition from the period of inflation \cite{infl}, 
during which the energy-momentum
tensor is dominated by the coherent inflaton field, to the radiation phase of Standard Big Bang
cosmology, is an important aspect of inflationary cosmology. Without such an energy transfer,
inflation would produce a cold empty universe and would not be a viable early universe scenario.
On the other hand, there will inevitably be gravitational particle production of any non-conformal
field which lives in the space-time of an inflationary universe \cite{Parker} (see \cite{BD} for a
classic review of quantum field theory in curved space-time). The energy density produced by
this mechanism by the end of the inflationary phase will be of the order $H^4$, where $H$ is
the Hubble expansion rate during inflation. Note that this energy scale is parametrically
suppressed compared to the energy density $\rho_I$ during inflation:
\be \label{supp}
\frac{H^4}{\rho_I} \, \sim \, \frac{H^2}{m_{pl}^2} \, ,
\ee
where $m_{pl}$ is the Planck mass. This ratio is bounded to be smaller than
$10^{-8}$ based on the upper bound on the strength of gravitational radiation
produced during inflation \cite{GWbound}.

In simple scalar field models of inflation, based on a slowly rolling scalar field with
canonical kinetic term in the action, there is a much more efficient energy
transfer mechanism which can reheat the universe. In the presence of any
coupling between the matter field (here modelled as a scalar field $\chi$) and
the inflaton field $\phi$ there is a parametric resonance instability which causes
$\chi$ field fluctuations to grow exponentially during the phase after inflation when
$\phi$ oscillates coherently about its ground state value \cite{TB, DK}. Although
this process does not directly produce a thermal state of matter particles, it efficiently
transfers the energy density from the inflaton to matter, typically in a period which
is short compared to a Hubble expansion time. This process is now called {\it preheating}
\cite{KLS1, STB, KLS2} (see \cite{RHrevs} for recent reviews). It produces a
state after inflation in which the matter energy density 
\be
\rho_m \, \sim \, \rho_I
\ee
after inflation is not suppressed compared to the energy density $\rho_I$ during
inflation, in contrast to what is obtained (\ref{supp}) if only gravitational
particle production is operative.

Simple slow-roll inflation based on an action with a canonical kinetic term 
is at the moment consistent with the data we have.
In fact, the scenario made a number of successful predictions (spatial flatness, slight
red tilt \cite{Mukh1} to the spectrum of cosmological perturbations, etc.) 
\cite{Guth, Linde, Mukh2} \footnote{But see \cite{Ijjas} for a different view.}. 
The scenario also predicts a red tilt in the spectrum of gravitational waves \cite{Starob0}.

There are alternatives to the inflationary paradigm of early universe
cosmology (see e.g. \cite{RHBrev} for reviews). One of these alternatives,
{\it String Gas Cosmology} \cite{BV} (see also \cite{Perlt}), while consistent
with all current observations of scalar cosmological perturbations \cite{NBV},
predicts a slight blue tilt in the spectrum of gravitational waves \cite{BNPV}.
In the context of inflationary cosmology with vacuum initial conditions and
with matter obeying the {\it Null Energy Condition} (NEC), one always obtains
a red tilt \footnote{This is different than the scalar spectrum for which either a
red or a blue tilt can be obtained, although the simplest slow-roll models of
inflation also predict a red tilt of the scalar spectrum.}.

However, it was pointed out in \cite{Yoko} (see also \cite{Vikman}), that by introducing Galileon
type terms (in particular kinetic terms) in the action of a scalar field $\phi$, it is possible 
to obtain an inflationary model in which matter violates the NEC and hence a blue
tensor tilt is possible \footnote{It is still possible to distinguish String
Gas Cosmology from G-Inflation by considering non-Gaussianities
or consistency relations \cite{Recent}.} . This model is called {\it G-inflation}. In this model,
inflation is driven by the kinetic term in the action which at early times has
the ``wrong'' sign and hence can lead to the violation of the NEC. 
Nevertheless thanks to the Galileon-type terms, the stability of fluctuations 
is maintained even in the presence of NEC violation contrary to the case of
k-inflation \cite{ArmendarizPicon:1999rj}.
Inflation
terminates at a scalar field value above which the sign of the kinetic term
reverts back to its canonical form. This leads to a transition from an inflationary
phase to a standard kinetic-driven phase with equation of state $w = 1$,
where $w$ is the ratio of pressure to energy density. The energy density
in $\phi$ then decreases as $a(t)^{-6}$, where $a(t)$ is the cosmological
scale factor. 

Since there is no phase during which $\phi(t)$ oscillates there 
is no possibility of preheating. As discussed in \cite{Yoko}, the
production of regular matter particles after Galileon inflation is still
possible by the gravitational Parker particle production mechanism.
However, the resulting matter energy density will be suppressed as
in (\ref{supp}). The question we ask in this note is whether in the
presence of a coupling between matter and the inflaton there is
nevertheless some non-gravitational channel which transfers
energy to matter faster than what can be achieved by gravitational
effects.

In the following we point out that there is indeed a channel for direct
particle production, and we derive conditions on the coupling constant
for which this direct channel is more efficient than Parker particle
production \footnote{A similar channel is operative in the ``emergent
Galileon'' scenario of \cite{Rubakov} - see \cite{Laurence}. Particle-induced
particle production has also recently been studied in a bouncing
cosmology in \cite{Jerome}.}. 
Our analysis is based on the general framework set out
in \cite{TB}.

We begin with a brief review of G-inflation, move on to a
discussion of the particle production mechanism we use, before
presenting the calculations applied to our model. We work in natural
units in which the speed of light, Planck's constant and Boltzmann's
constant are set to $1$.

\section{G-inflation}

The original G-inflation \cite{Yoko} is based on a scalar field $\phi$
minimally coupled to gravity with an action
\be
{\cal L} \, = \, K(\phi, X) - G(\phi, X) \square \phi \, ,
\ee
where $X$ is the standard kinetic operator
\be
X \, = \, \frac{1}{2} \partial_{\mu} \phi \partial^{\mu} \phi \, ,
\ee
and $K$ and $G$ are general functions of $\phi$ and $X$. 
See \cite{Kobayashi:2011nu} for its generalized version.
The special property
of this class of Lagrangians is that the resulting equations of
motion contain no higher derivative terms than second order \cite{higher}.  In the
case $K = K(X)$ and $G(\phi, X) \propto X$ the action has an extra 
shift symmetry (``Galilean symmetry'') and these Lagrangians
were introduced and studied in \cite{Galileon}. 

The model of kinetically driven G-inflation \cite{Yoko} is based on
choosing
\be
K(\phi, X) \, = \, - A(\phi) X + \delta K \, ,
\ee
with
\be
A(\phi) \, = \, {\rm tanh} \bigl[ \lambda (\phi_e - \phi) \bigr] \, ,
\ee
and 
\be
G(\phi, X) \, = \, {\tilde{g}}(\phi) X \, = \, {\tilde{g}} X \, .
\ee
Here $\lambda$ and ${\tilde{g}}$ are coupling constants and $\delta K$
includes higher order terms in $X$ which are  important during inflation.  
After the sign of the linear kinetic term in the action is flipped at 
 $\phi = \phi_e$, they soon become negligible and do not affect our analysis
 of reheating.

We consider homogeneous and isotropic cosmological solutions
resulting from this action.
As shown in \cite{Yoko}, for $\phi < \phi_e$ there are inflationary
trajectories for which the quasi-exponential expansion of space is
driven by the wrong-sign kinetic term. Inflation ends at $\phi = \phi_e$,
and for $\phi > \phi_e$ the background becomes that of a
kinetic-driven phase with $w = 1$, $a(t) \sim t^{1/3}$ and
\be \label{scaling}
{\dot \phi}(t) \, \sim \, \frac{1}{t} \, .
\ee
We call this stage the {\it kination regime} of the model. 
Since the energy density in $\phi$ decays so rapidly, eventually
the kination regime will end and regular radiation and matter will begin to dominate.
The energy density at which this transition happens determines the
{\it reheating temperature} of the Universe. 

Knowledge of the reheating temperature is important for various post-inflationary
processes such as baryogenesis or the possibility of production of
topological defects. It may also be possible to directly probe the
physics of the phase between the initial thermal stage and the
hot Big Bang phase with precision observations (see e.g. \cite{Kamion}).

Regular matter and radiation are produced by gravitational particle
production. However, if this is the only mechanism, then the
reheating temperature will be low as it is suppressed by (\ref{supp}).
In the following we will assume that there is a direct coupling
between matter (described by a free massless scalar field $\chi$) and the
inflaton field $\phi$. We consider two possible couplings. The first is of the form
\be \label{intlag}
{\cal L}_I \, = \, \frac{1}{2} g^2 {\dot{\phi}} \chi^2 \, ,
\ee
where $g$ is a dimensionless coupling constant. Note
that we have chosen a derivative coupling of $\phi$
with $\chi$ to preserve the invariance of the interaction
Lagrangian under shifting of the value of $\phi$ (which
is part of the Galilean symmetry. The disadvantage of this
coupling is that it violates the symmetry $\phi \rightarrow - \phi$.
The second coupling obeys this symmetry but involves 
non-renormalizable interactions. It is
\be \label{intlag2}
{\cal L}_I \, = \,  - \frac{1}{2} M^{-2} {\dot{\phi}}^2 \chi^2 \, ,
\ee
where $M$ is a new mass scale which is expected to be
smaller than the Planck mass. 
These couplings open up non-gravitational 
channels for the production of $\chi$ particles. In the following we 
will study the conditions under which these direct production channels 
are more efficient than the gravitational particle production channel.

\section{Inflaton-Driven Particle Production}

Assuming that the Lagrangian for the matter field $\chi$ has
canonical kinetic term, then the Lagrangian for $\chi$ is
that of a free scalar field with a time dependent mass, the
time dependence being given by the interaction Lagrangians (\ref{intlag})
or (\ref{intlag2}). Each Fourier mode $\chi_k$ of $\chi$ evolves independently,
the equation of motion is
\be \label{EoM1}
{\ddot{\chi}}_k + 3 H {\dot{\chi}}_k + \lmk \frac{k^2}{a^2} - g^2 {\dot{\phi}} \rmk \chi_k 
\, = \, 0 \, .
\ee
or
\be \label{EoM1b}
{\ddot{\chi}}_k + 3 H {\dot{\chi}}_k + \lmk \frac{k^2}{a^2} + M^{-2} {\dot{\phi}}^2 \rmk \chi_k 
\, = \, 0 \, ,
\ee
depending on the form of the interaction Lagrangian.
The effects of the expansion of space can be pulled out by
rescaling the field
\be
X_k \, \equiv \, a^{-1} \chi_k \, .
\ee
Then, in terms of conformal time $\tau$ (which is related to physical
time $t$ by $dt = a(t) d\tau$), the equation of motion becomes
\be \label{EoM2}
X_k^{\prime \prime}  + \lmk k^2 - g^2 {\dot{\phi}} a^2 - \frac{a^{\prime \prime}}{a} \rmk X_k \, = \, 0 \, ,
\ee
or
\be \label{EoM2b}
X_k^{\prime \prime}  + \lmk k^2 + M^{-2} {\dot{\phi}}^2 a^2 - \frac{a^{\prime \prime}}{a} \rmk X_k \, = \, 0 \, ,
\ee
where a prime denotes a derivative with respect to $\tau$.

The qualitative features of the equations of motion (\ref{EoM2}) or (\ref{EoM2b})
are well known from the theory of cosmological perturbations
(see e.g. \cite{MFB} for an in-depth review and \cite{RHBpertrev}
for a brief overview): In the absence of the interaction term,
$X_k$ will oscillate on sub-Hubble scales, i.e. scales for which
\be
k^2 \, > \, \frac{a^{\prime \prime}}{a} \, \sim {\cal H}^2 \, ,
\ee
whereas the mode function $X_k$ is squeezed on super-Hubble
scales, i.e. 
\be
X_k \, \sim \, a \, .
\ee

Following \cite{TB}, we will treat the effects of the interaction term
in leading order Born approximation, i.e. we write
\be
X \, \equiv \, X_0 + X_1 \, ,
\ee
(here and in the following we will drop the subscript $k$) where $X_0$ is
the solution of the equation in the absence of interactions, i.e. a solution of
\be
X_0^{\prime \prime} + \lmk k^2 - \frac{a^{\prime \prime}}{a} \rmk X_0 \, = \, 0 \, ,
\ee
solving the initial conditions of the problem,
and $X_1$ is the solution of the inhomogeneous equation 
\be \label{EoM3}
X_1^{\prime \prime} + \lmk k^2 - \frac{a^{\prime \prime}}{a} \rmk X_1 \, = \, g^2 {\dot{\phi}} a^2 X_0 
\ee
or
\be \label{EoM3b}
X_1^{\prime \prime} + \lmk k^2 - \frac{a^{\prime \prime}}{a} \rmk X_1 \, = \, - M^{-2} {\dot{\phi}}^2 a^2 X_0 
\ee
(with vanishing initial conditions)
obtained by taking the interaction term in (\ref{EoM2}) or (\ref{EoM2b}) to the right hand side
of the equation and replacing $X$ by the ``unperturbed'' solution $X_0$.

The inhomogeneous equation (\ref{EoM3}) (or (\ref{EoM3b})) can be solved by 
the Green's function method
\be
X_1(\tau) \, = \, \int_{\tau_i}^{\tau} d \tau' G(\tau, \tau') g^2 a^2(\tau') {\dot{\phi}}(\tau') X_0(\tau') \, ,
\ee
or
\be
X_1(\tau) \, = \, - \int_{\tau_i}^{\tau} d \tau' G(\tau, \tau') M^{-2} a^2(\tau') {\dot{\phi}}^2(\tau') X_0(\tau') \, ,
\ee
where the Green's function $G(\tau, \tau')$ is determined in terms of the two fundamental
solutions $u_1$ and $u_2$ of the homogeneous equation via
\be
G(\tau, \tau') \, = \, W^{-1} \bigl( u_1(\tau) u_2(\tau') - u_2(\tau) u_1(\tau') \bigr) \, ,
\ee
where $W$ is the Wronskian
\be
W \, = \, u_1(\tau) u_2^{\prime}(\tau) - u_2(\tau) u_1^{\prime}(\tau) \, .
\ee
In the above, the time $\tau_i$ is the time when the initial conditions are
imposed. In our case it is the end of the period of inflation.

The condition that direct particle production is more efficient than gravitational
particle production is
\be \label{crit}
X_1(\tau) \, > \, X_0(\tau)
\ee
at some time $\tau > \tau_i$ before the time when the kinetic phase would be
terminated by gravitational particle production alone.

\section{Analysis}

We now apply the formalism of the previous section to our specific
Galileon inflation model. We are interested in super-Hubble modes for
which the $k^2$ term in the equation of motion (\ref{EoM1}) can be
neglected. The fundamental solutions are then
\ba
u_1(\tau) \, &=& \, \lmk \frac{\tau}{\tau_i} \rmk^{1/2}\\
u_2(\tau) \, &=& \, \lmk \frac{\tau}{\tau_i} \rmk^{1/2} {\rm ln}\lmk\frac{\tau}{\tau_i}\rmk \, , \nonumber
\ea
and hence the Wronskian becomes
\be
W \, = \, \frac{1}{\tau_i} \, ,
\ee
and the Green's function is
\be
G(\tau, \tau') \, = \bigl( \tau \tau' \bigr)^{1/2} {\rm ln}\lmk\frac{\tau'}{\tau}\rmk \, .
\ee

The contribution $X_1(\tau)$ induced by the direct coupling between
$\phi$ and $\chi$ thus becomes
\be
X_1(\tau) \, = \, g^2 \int_{\tau_i}^{\tau} d\tau' \bigl( \tau \tau' \bigr)^{1/2} {\rm ln} \lmk\frac{\tau'}{\tau}\rmk
{\dot{\phi}}(\tau') a^2(\tau') X_0(\tau') \, ,
\ee
or
\be
X_1(\tau) \, = \, \frac{-1}{M^{2}} \int_{\tau_i}^{\tau} d\tau' \lmk \tau \tau' \rmk^{1/2} {\rm ln} \lmk\frac{\tau'}{\tau}\rmk
{\dot{\phi}}(\tau')^2 a^2(\tau') X_0(\tau') \, ,
\ee
For $X_0(\tau)$ we can take the dominant solution of the homogeneous
equation
\be
X_0(\tau') \, = \, X_0(\tau_i) \lmk \frac{\tau'}{\tau_i} \rmk^{1/2} {\rm ln}\lmk\frac{\tau'}{\tau_i}\rmk \, .
\ee
Making use of the scaling (\ref{scaling}) of ${\dot{\phi}}$ and after a couple of lines of
algebra we obtain the approximate result (keeping only the contribution from the upper
integration limit)
\be
X_1(\tau) \, \simeq \, g^{2} {\dot{\phi_i}}(\tau_i) \tau^2 X_0(\tau_i) \, .
\ee
or
\be
X_1(\tau) \, \simeq \, -M^{-2} {\dot{\phi_i}}^2(\tau_i) \tau_i^2 \lmk \frac{\tau}{\tau_i} \rmk^{1/2} X_0(\tau_i) 
{\rm ln} \lmk\frac{\tau}{\tau_i} \rmk\, .
\ee
If we take the initial time $\tau_i$ to correspond to the end of inflation, we
have
\be
{\dot{\phi}}(\tau_i) \, \simeq \, H(\tau_i) m_{pl} \, ,
\ee
where $H(\tau_i)$ is the value of $H$ at the end of inflation. In this case
\be
X_1(\tau) \, \sim \, g^2 m_{pl} \tau_i \lmk \frac{\tau}{\tau_i} \rmk^{3/2} \, ,
\ee
or
\be
X_1(\tau) \, \simeq \, - \lmk \frac{m_{pl}}{M} \rmk^2 X_0(\tau) \, .
\ee

The criterion (\ref{crit}) for direct particle production to dominate over
gravitational particle production then becomes (up to logarithmic factors)
\be
g^2 \, > \, \frac{H(\tau_i)}{m_{pl}} \lmk \frac{t_i}{t} \rmk \, .
\ee
or
\be
\lmk \frac{m_{pl}}{M} \rmk^2  \, > \, 1 \, .
\ee
Note that for the second interaction term, particle production via
direct interactions dominates within one Hubble expansion time (the
time interval after which the contribution from the lower integration
end can be neglected), provided that $M < m_{pl}$, a condition
which has to be satisfied if we are to trust the effective field justification
of the interaction term.

Once $X_1(\tau)$ starts to dominate over $X_0(\tau)$, the Born
approximation ceases to be valid. At that point, the coupling term
in the equation of motion for $X$ will become the dominant one,
and an approximation to (\ref{EoM2}) (we will first focus on the
case of the first interaction term) which is self-consistent for
long wavelength modes (for which the $k^2$ term in the equation
is negligible) is
\be \label{EoM4}
X^{\prime \prime} - g^2 {\dot{\phi}} a^2 X \, = \, 0 \, .
\ee
An approximate solution of this equation is
\be \label{ansatz}
X(\tau) \, = \, {\cal{A}}(\tau) e^{{\tilde{f}} (\frac{\tau}{\tau_i})^{3/4} \tau_i}
\ee
with
\be \label{ansatz2}
{\tilde{f}} \, \equiv \, \frac{4}{3} \bigl( g^2 {\dot{\phi}}(\tau_i) \bigr)^{1/2} \, .
\ee
Inserting this ansatz (\ref{ansatz} and \ref{ansatz2}) into (\ref{EoM4})
we find an equation for the amplitude ${\cal{A}}(\tau)$
\be
{\cal{A}}^{\prime \prime} + \frac{3}{2} {\tilde{f}} \tau^{- 1/4} \tau_i^{- 3/4} {\cal{A}}^{\prime}
- \frac{3}{16} {\tilde{f}} \tau^{-5/4} \tau_i^{- 3/4} {\cal{A}} \, = \, 0 \, ,
\ee
which both for ${\tilde{f}} \tau_i \ll 1$ and ${\tilde{f}} \tau_i \gg 1$ has a dominant
solution which is constant in time.

From (\ref{ansatz}) and (\ref{ansatz2}) we see that there is quasi-exponential
growth of $X$ which becomes important once
\be
{\tilde{f}} \tau_i \lmk \frac{\tau}{\tau_i} \rmk^{3/4} \, > \, 1 \, ,
\ee
which in terms of physical time is
\be \label{bound}
\frac{t}{t_i} \, > \, {\tilde{f}}^{-2} \tau_i^{-2} \, .
\ee
In the above we are implicitly assuming that ${\tilde{f}} \tau_i < 1$. If
${\tilde{f}} \tau_i > 1$ then reheating via direct particle production is
instantaneous on a Hubble time scale and the reheating temperature
is given by the energy density at the end of inflation.

Returning to the case ${\tilde{f}} \tau_i  < 1$, the we see that
once the time $t$ is larger than the one given by (\ref{bound}), the energy
transfer from the inflaton to matter is exponentially fast and will
immediately drain all of the energy from the inflaton. Hence, the
``reheating time'' $t_{RH}$ is
\be
t_{RH} \, \sim \, t_i ({\tilde{f}} \tau_i)^{-2} \, ,
\ee
and since the energy density between $t_i$ and $t_{RH}$
decreases as $a(t)^{-6} \sim t^{-2}$ we have
\be
\rho(t_{RH}) \, \sim \, \rho(t_i) ({\tilde{f}} \tau_i)^4 \, .
\ee
Making use of $\rho(t_i) = H^2(t_i) m_{pl}^2$ (up to a numerical factor)
and $\rho(t_{RH}) \sim T_{RH}^4$ we finally obtain the reheating
temperature $T_{RH}$ to be
\be
T_{RH} \, \sim \, {\tilde{f}} \tau_i ( H(t_i) m_{pl} )^{1/2} \, 
\ee
which is larger than the reheating temperature $H(t_i)$ which would
be obtained if only gravitational particle production were effective, provided that
\be
{\tilde{f}} \tau_i \, > \lmk \frac{H(t_i)}{m_{pl}} \rmk^{1/2} \, .
\ee

In the case of the second coupling, the conclusions are similar. Once the
coupling term in the equation of motion dominates over the expansion term,
the equation can be approximated as (changing the sign of the coupling term)
\be
X^{\prime \prime} - M^{-2} {\dot{\phi}}^2 a^2 X \, = \, 0 \, .
\ee
Since
\be
{\dot{\phi}}^2(\tau) a^2(\tau) \, = \, {\dot{\phi}}^2(\tau_i) \lmk \frac{\tau_i}{\tau} \rmk^2 \, ,
\ee
the equation has power law solutions with an exponent $\Delta$ given by
\be
\Delta \, = \, \frac{1}{2} \bigl[ 1 \pm \sqrt{1 + 4 R^2} \bigr] \, ,
\ee
where
\be
R \, \equiv \, \frac{m_{pl}}{M} \, .
\ee
We see that if $M \ll m_{pl}$, then the power of the dominant solution is $\Delta \gg 1$
and this means that there is complete energy transfer from the inflaton to $\chi$
within one Hubble expansion time. Hence, the reheating temperature is given
by the energy density at the end of inflation, i.e.
\be
T_{RH} \, \sim \, ( H(t_i) m_{pl} )^{1/2} \, .
\ee

\section{Conclusions and Discussion}

We have derived the condition under which direct particle production in
G-inflation dominates over gravitational particle production. 
The discussion also applies to k-inflation \cite{ArmendarizPicon:1999rj}.
We consider
two possible interaction Lagrangians, namely (\ref{intlag}) and (\ref{intlag2}). 
We first study the onset of matter particle production from the direct coupling
using the Born approximation. We find that for both interaction terms we
consider the direct particle production channel eventually dominates.
This happens within one Hubble expansion time for the coupling (\ref{intlag2}),
whereas in the case of (\ref{intlag}) the time when direct particle production
starts to dominate depends on the coupling constant $g$. 

Once direct particle production begins to dominate over gravitational particle
production we must use a different approximation scheme to solve the
equation of motion. We can now neglect the squeezing term in the equation
of motion. We provide solutions of the resulting approximate equations
of motion and show that once direct particle production begins to dominate,
the energy transfer from the inflaton to the matter fields will be almost
instantaneous. This allows us to estimate the value of
the reheating temperature, the temperature of matter once the inflaton field
has lost most of its energy density to particle production. In the case of
the second interaction term (\ref{intlag2}), the reheating temperature is
given by the energy density at the end of inflation, in the case of the
first interaction term (\ref{intlag}), it is reduced by a factor which involves
the interaction coupling constant $g$.

In the present work we have studied the production of particles which
correspond to an entropy fluctuation direction. There is the danger 
(see e.g. \cite{Fabio2} for an initial study and \cite{Hossein} for
recent work) that the induced entropy fluctuations might induce
too large curvature perturbations, as it does in the model studied
in \cite{Kunimitsu}. A study of this question will be the focus of
future work.
\\

\section*{Acknowledgements}

One of us (RB) wishes to thank the Institute for Theoretical Studies of the ETH
Z\"urich for kind hospitality during the 2015/2016 academic year. 
RB acknowledges financial support from Dr. Max
R\"ossler, the  ``Walter Haefner Foundation'' and the ``ETH Zurich Foundation'', and
from a Simons Foundation fellowship. The research is also supported in
part by funds from NSERC and the Canada Research Chair program. 
HBM also wishes to thank the Institute for Theoretical Studies of the ETH
Z\"urich for kind hospitality while this work was being prepared. 
HBM is supported in part by an Iranian MSRT fellowship. 
This work was partially supported by Open Partnership Joint Projects Grant of
JSPS.

\section*{Appendix}

In this Appendix we compare our calculation of particle production in
G-inflation with what is obtained using more standard methods
of quantum field theory in curved space-time (see e.g. \cite{Parker, Ford, Starob}).
Recall the equation of motion for our canonical field $X$. In the case
of the second coupling which we consider in the main text this equation is
\be \label{EoM2c}
X_k^{\prime \prime}  + \lmk k^2 + M^{-2} {\dot{\phi}}^2 a^2 - \frac{a^{\prime \prime}}{a} \rmk X_k \, = \, 0 \, .
\ee
We will compare the energy density in produced particles in the initial stages
of particle production (before back-reaction becomes important).

Starting from vacuum initial conditions, we can obtain the energy density
from the Bogoliubov coefficients $\beta_k$ which describe how the
solution of (\ref{EoM2c}) that initially corresponds to the vacuum can
at late times $\tau$ be decomposed into positive and negative frequency modes
(see e.g. \cite{BD} for a textbook treatment):
\be \label{endens}
\rho(\tau) \, = \, \frac{1}{2 \pi^2 a^4(\tau)} \int_0^{\infty} |\beta_k|^2 k^3 dk \, .
\ee
The Bogoliubov coefficient $\beta_k$ is given by
\be
\beta_k \, = \, \frac{i}{2k} \int_{-\infty}^{\infty} e^{-2ik\tau} V(\tau) d\tau \, ,
\ee
where in the case of the equation (\ref{EoM2c}) 
\be
V(\tau) \, = \,  \lmk12 \frac{m_{pl}^2}{M^2}-2\rmk\frac{1}{\tau^2} \, 
\ee
(for times in the kination phase).
The first term on the right hand side of this equation represents particle
production via the particle interactions, whereas the second term corresponds
to gravitational particle production. From this equation it is already
clear that for $M \ll m_{pl}$ particle interactions will dominate the energy transfer
from the inflaton field to matter.

Following \cite{Kunimitsu}, the expression (\ref{endens}) for the energy density
can be rewritten as
\be \label{endens2}
\rho(\tau) \, = \, \frac{-1}{32 \pi^2 a^4} \int_{-\infty}^{\tau} d\tau_1 d\tau_2 
\ln(\mu\left|(\tau_1 - \tau_2)\right|) V'(\tau_1)V'(\tau_2) \,
\ee
where $\mu$ is a regularization scale which has been introduced
to remove the ultraviolet divergence. Note that the derivative of $V$
in the above equation is with respect to conformal time. 

If we are
interested in particle production due to the squeezing of the mode
wave functions on super-Hubble scales, we have the natural
regularization scale $\mu = H(\tau_i)$, the Hubble scale at the end
of inflation. Making use of this cutoff, it can be shown that the
order of magnitude of $\rho(\tau)$ obtained from (\ref{endens2})
agrees with what is obtained using the method we have used in
the main text, namely
\be
\beta_k(\tau) \, \sim \, \frac{X_k(\tau)}{X_k^{vac}} \, 
\ee
where $X_k^{vac}$ is the vacuum value of the canonical variable.

\end{document}